# DEGRADATION OF 2-MERCAPTOBENZOTHIZAOLE IN MICROBIAL ELECTROLYSIS CELLS: INTERMEDIATES, TOXICITY, AND MICROBIAL COMMUNITIES.

M. Isabel San-Martín[a], Adrián Escapa[a,c*], Raúl M. Alonso[a], Moisés Canle[b], Antonio Morán[a]

[a] Chemical and Environmental Bioprocess Engineering Group, Natural Resources Institute (IRENA), Universidad de León, Avda. de Portugal 41, Leon E-24009, Spain.

[b] Chemical Reactivity & Photoreactivity Group, Dept. of Chemistry, Faculty of Sciences & CICA, University of A Coruña. E-15071 A Coruña, Spain.

[c] Department of Electrical Engineering and Automatic Systems, Universidad de León, Campus de Vegazana s/n, E-24071 León, Spain.

* Corresponding author:

Tel.: 0034 987293378

E-mail: adrian.escapa@unileon.es

## Highlights

- Five degradation routes and two additional dimerization routes were identified
- The treated electrodes promoted the presence of enriched and specific communities
- Graphene electrodeposition on the anode favours the mercaptobenzothiazole degradation
- A microbial electrolysis cell can reduce biotoxicity from 46 to 28 eqtox·$m^{-3}$

# Abstract

The compound 2-mercaptobenzothizaole (MBT) has been frequently detected in wastewater and surface water and is a potential threat to both aquatic organisms and human health (its mutagenic potential has been demonstrated). This study investigated the degradation routes of MBT in the anode of a microbial electrolysis cell (MEC) and the involved microbial communities. The results indicated that graphene-modified anodes promoted the presence of more enriched, developed, and specific communities compared to bare anodes. Moreover, consecutive additions of the OH substituent to the benzene ring of MBT were only detected in the reactor equipped with the graphene-treated electrode. Both phenomena, together with the application of an external voltage, may be related to the larger reduction of biotoxicity observed in the MEC equipped with graphene-modified anodes (46.2 eqtox·m$^{-3}$ to 27.9 eqtox·m$^{-3}$).

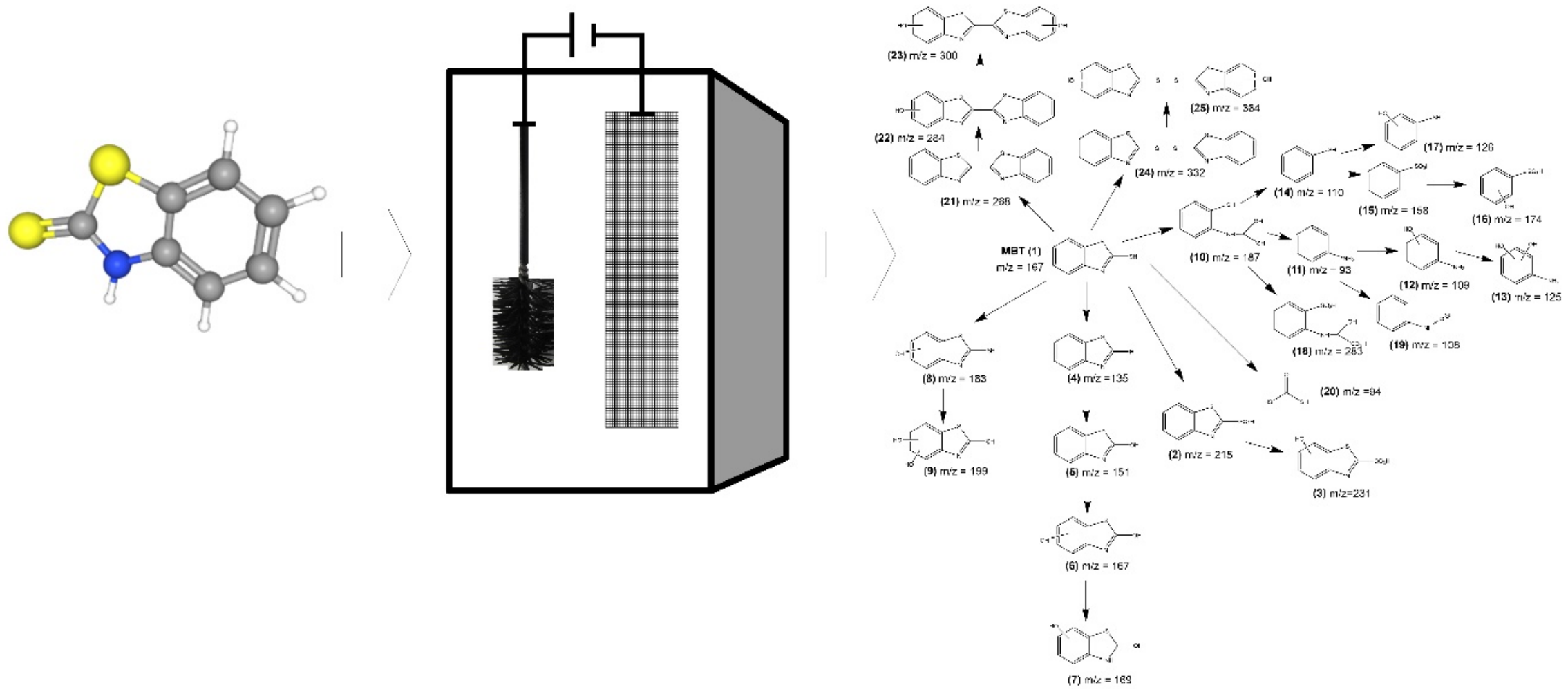

# 1. INTRODUCTION

The benzothiazoles (BTHs) are a group of heterocyclic aromatic compounds among which 2-mercaptobenzothiazole (MBT) is the most important member (Gaja and Knapp, 1998). This xenobiotic compound is used mainly as a corrosion inhibitor and can be applied for the synthesis of antibiotics, herbicides, pesticides, rubber, or leather (Li et al., 2004). It is a toxic, widespread, and poorly biodegradable pollutant of the aquatic environment, and many studies have confirmed that it is a strong allergen and potential mutagen for humans (Morsi et al., 2020a)(Allaoui et al., 2010). However, conventional biological wastewater treatment methods cannot effectively remove MBT from waste streams (Derco et al., 2014) because of the difficulty to biodegrade it (De Wever and Verachtert, 1994). As a consequence, the development of alternative technologies, such as photocatalytic degradation (Li et al., 2005) or the Fenton reaction (Wang et al., 2016), to mitigate its presence in contaminated streams has aroused extensive interest.

Bioelectrochemical systems (BESs) comprise a group of bio-based technologies that have considerable potential for wastewater treatment (Hua et al., 2019) and for bioremediation of a wide variety of organic and inorganic pollutants (Wang et al., 2020), including xenobiotics (Fernando et al., 2019). For BTH, and to our knowledge, only one brief study reported its degradation in a BES (Liu et al., 2016). In the referred work, the authors showed that the BES could effectively remove BTH from a simulated waste stream containing BTH in concentrations from 20 to 110 $mg \cdot L^{-1}$. However, neither the BTH degradation pathways nor the involved microbial communities (and their role) were presented. In our study we aim precisely at closing this gap by (i) shedding light on the degradation routes of MBT in the anode of a BES, (ii) identifying the microbial communities that may be related to MBT degradation, and (iii) assessing to what extent modifying the anode surface can reduce the biotoxicity of MBT-containing effluents.

As the central working principle of BES relies on the exchange of electrons between electroactive microorganisms and solid electrodes, the optimization of this interaction has the potential to significantly improve the performance of these systems (Guo et al., 2015). In this regard, the utilization of nanomaterials like carbon nanotubes (CNTs), metal nanoparticles or graphene oxide (GO) offers new perspectives (Morsi et al., 2020b) due to their unique characteristics of chemical stability, high conductivity and enhanced electro-catalytic activity for a variety of redox-reactions (ElMekawy et al., 2017). In this study GO was used as a modifying agent because of its recently explored ability to promote the degradation of complex organic compounds by acting synergistically with microorganisms (Colunga et al., 2015; Khalid et al., 2018; Shen et al., 2018). Another aspect that makes GO interesting as an electrode modification agent is its relatively low cost, availability and the possibility of using non-polluting procedures for its reduction (Chua and Pumera, 2014).

# 2. MATERIAL AND METHODS

## 2.1 MEC setup and operation

The experiments were conducted in duplicates in the BES operated as microbial electrolysis cells (MECs). Applied voltage was set at 1V (this value is high enough to overcome electrodes overpotentials and low enough to avoid water electrolysis). MBT degradation in the anode of these cells was studied under three conditions (thus, six cells were used). The first condition involved the use of graphene-modified electrodes and was conducted in two MECs: MEC-GO1 and MEC-GO2. Graphene oxide (GO) was electrodeposited on carbon brush electrodes through a series of 16 consecutive cyclic voltammetries between −1.5 and 0.8 V vs. Ag/AgCl (3 M) at a scan rate of 20 $mVs^{-1}$ by following the method described in Alonso et al. (2017). The cyclic voltammograms obtained during the electrodeposition cycles are shown in the supporting information (Figure A.1). To confirm that GO was successfully electrodeposited on the electrode, SEM analyses of the modified and unmodified electrodes were performed (Figure A2). For the second condition, bare (unmodified) carbon brush anodes were used in two MECs that were periodically operated under open-circuit (OC) conditions. These MECs were labelled as MEC-OC1 and MEC-OC2. Finally, two MECs were used as controls (labelled as MEC-UN1 and MEC-UN2) in which bare (untreated) carbon brush anodes were used.

The six MECs were constructed with modified 250-mL Duran® bottles. The anode and cathode were made of a carbon brush and stainless steel mesh, respectively. The carbon brush consisted of carbon fibres (Mill-Rose, USA) distributed in a twisted titanium wire backbone with a length of 3 cm and an outer diameter of 2 cm. The stainless steel mesh had a projected area of 15 $cm^2$ (6 x 2.5 cm).

River mud (obtained from Porma river) was used as the source of inoculum for all MECS and was diluted in culture media (1:4 dilution rate) before being fed to the anodic chambers of the MECs. After inoculation, the reactor was fed with culture media and with 0.55 $g \cdot L^{-1}$ sodium acetate as the carbon source. The culture media contained (in $g \cdot L^{-1}$) $NH_4Cl$ at 0.15, $NaHCO_3$ at 0.1, NaCl at 0.5, $MgSO_4$ at 0.015, $CaCl_2$ at 0.02, $K_2HPO_4$ at 1.07, $KH_2PO_4$ at 0.53, a trace mineral solution at 10 mL, and vitamins at 1 mL, as reported by del Pilar Anzola Rojas et al. (2018), added to demineralized water. After acclimation, MBT was added to the synthetic solution at a concentration of 0.05 mM, and a subsequent stirring for 6 h and purging with $N_2$ gas for 15 min were performed before placing it into the reactor.

All operations were conducted in fed-batch mode (2–3 days) at room temperature (21 ± 2 °C). The voltage to each MEC was applied using an EA 2042-06 power supply (Elektro-automatik, German). The electrical current was calculated by measuring the voltage drop through a 10 Ω resistor and using Ohm's law ($I = U \cdot R^{-1}$). The voltage was measured and recorded at 10-min intervals using a Keithley 2700 multimeter (Keithley Instruments, USA).

## 2.2 Chemical analysis and calculation

The total organic carbon (TOC) and pH of the medium were measured at the beginning and the end of each batch cycle using a TOC multi N/C 3100 (Analytikjena, Germany) and a pH meter, GLP 21 (Crison Instruments, Spain), respectively. MBT and its transformation products were identified at different times (0, 7.5, 15, 30, and 50 h) during the last batch cycle by HPLC–MS, using a Thermo Scientific LTQ Orbitrap Discovery apparatus equipped with an electrospray interface operating both in positive ion mode (ESI+) and negative ion mode (ESI−). A Phenomenex Kinetex XB-C18 (100 mm × 2.10 mm, 2.6 µm) column was used. Analyses were carried out using full-scan data-dependent MS scanning from m/z 50 to 400. The structures of the transformation products were proposed by interpreting their corresponding MS spectra.

Coulumbic efficiency was calculated according to eq. 1 as:

$$Coulombic\ efficiency = \frac{\int_0^t I\,dt}{(COD_{in} - COD_{out})/M \cdot Q \cdot e \cdot F} \quad (1)$$

Where $COD_{in}$ and $COD_{out}$ are the COD concentration of BES influent and effluent, respectively, I is the circulating electrical current (A), M is the weight of 1 mol of COD (32 $g{\cdot}mol^{-1}$), Q is the influent flow rate ($L{\cdot}d^{-1}$), e is the number of mol of electrons exchanged per mol of COD equivalent consumed (4 $mol{\cdot}mol^{-1}$), and F is the Faraday constant (96,485 $C{\cdot}mol^{-1}$). TOC was converted to COD considering sodium acetate as the sole carbon source.

## 2.3. Biotoxicity analysis

Biotoxicity was assessed by determining the luminescence inhibition of the marine Gram-negative bacterium, *Vibrio fischeri* strain NRRL B-11177 (formerly *Photobacterium phosphoreum*), after 15 and 30 min of contact time. The bacteria were purchased as liquid dried kits, which were stored in a freezer at −20°C and rehydrated with liquid medium prior to testing. The light emission of this bacterium when in contact with different samples and exposure times was measured using a Microtox® 500 (Modern Water, UK) analyser. All samples were tested in triplicate.

## 2.4. Microbial community analysis

Genomic DNA was extracted for the carbon brush electrode using the PowerSoil® DNA Isolation Kit (MoBio Laboratories Inc., Carlsbad, CA, USA), following the manufacturer's instructions. The entire DNA extract was used for pyrosequencing of the eubacteria 16S-rRNA gene-based massive library. The primer set used was 27Fmod (5`-AGRGTTTGATCMTGGCTCAG-3`) /519R modBio (5`-GTNTTACNGCGGCKGCTG-3`), following the method described in San-Martín et al. (2019).

The DNA reads were compiled in FASTq files for further bioinformatics processing. The following steps were performed using QIIME. Denoising was performed using a denoiser (Chen et al., 2010). OTUs were then taxonomically classified using the Ribosomal Database Project (RDP) Bayesian Classifier (http://rdp.cme.msu.edu) and compiled into each taxonomic level with a bootstrap cut off value of 80% (Cole et al., 2009). Raw pyrosequencing data obtained from this analysis were deposited in the Sequence Read Archive of the National Centre for Biotechnology Information.

The quantitative analysis of the Eubacteria population was performed via a quantitative-PCR reaction (qPCR) using the PowerUp SYBR Green Master Mix in a StepOne plus Real Time PCR System (Applied Biosystems, USA). The primer sets were 518R qPCR (5'-ATTACCGCGGCTGCTGG-3') and 314F qPCR (5'-CCTACGGGAGGCAGCAG-3).

# 3. RESULTS AND DISCUSSION

## 3.1. MEC performance

The six MECs were all inoculated following the same procedure (i.e. identical inoculum, culture medium, temperature, and pH), and the current started to increase ~5 days after inoculation.

To favour the adaptation of microorganisms to the anodic environment and promote repeatable state conditions, all MECs were operated in batch mode from day 15 until day 51. The duration of each batch cycle (determined by the moment in which the current fell below 0.1 mA) was ~2 days, thus resulting in 18 batch cycles. The current (averaged for the two MECs used for each of the three conditions described in Section 2.1) is shown in Figure 1.

The currents in the MEC-GO reactors (MEC-GO1 and MEC-GO2) were slightly higher than in the MEC-UN (MEC-UN1 and MEC-UN2) and MEC-OC reactors (MEC-OC1 and MEC-OC2). This was especially visible during the first three cycles (Figure 1), where the peak currents were on average 23% and 38% higher compared to those for MEC-UN and MEC-OC, respectively. Afterwards, the current profiles tended to converge, which is in agreement with our earlier observations (Alonso et al., 2017), where graphene-treated electrodes provided relatively high current densities for the first few days of operation. The periodic power interruptions to which the MEC-OC cells were subjected did not seem to affect the performance in the cells, as current production recovered almost immediately once the electrical power was restored (Figure 1). A similar conclusion was reached by del Pilar Anzola Rojas et al. (2018).

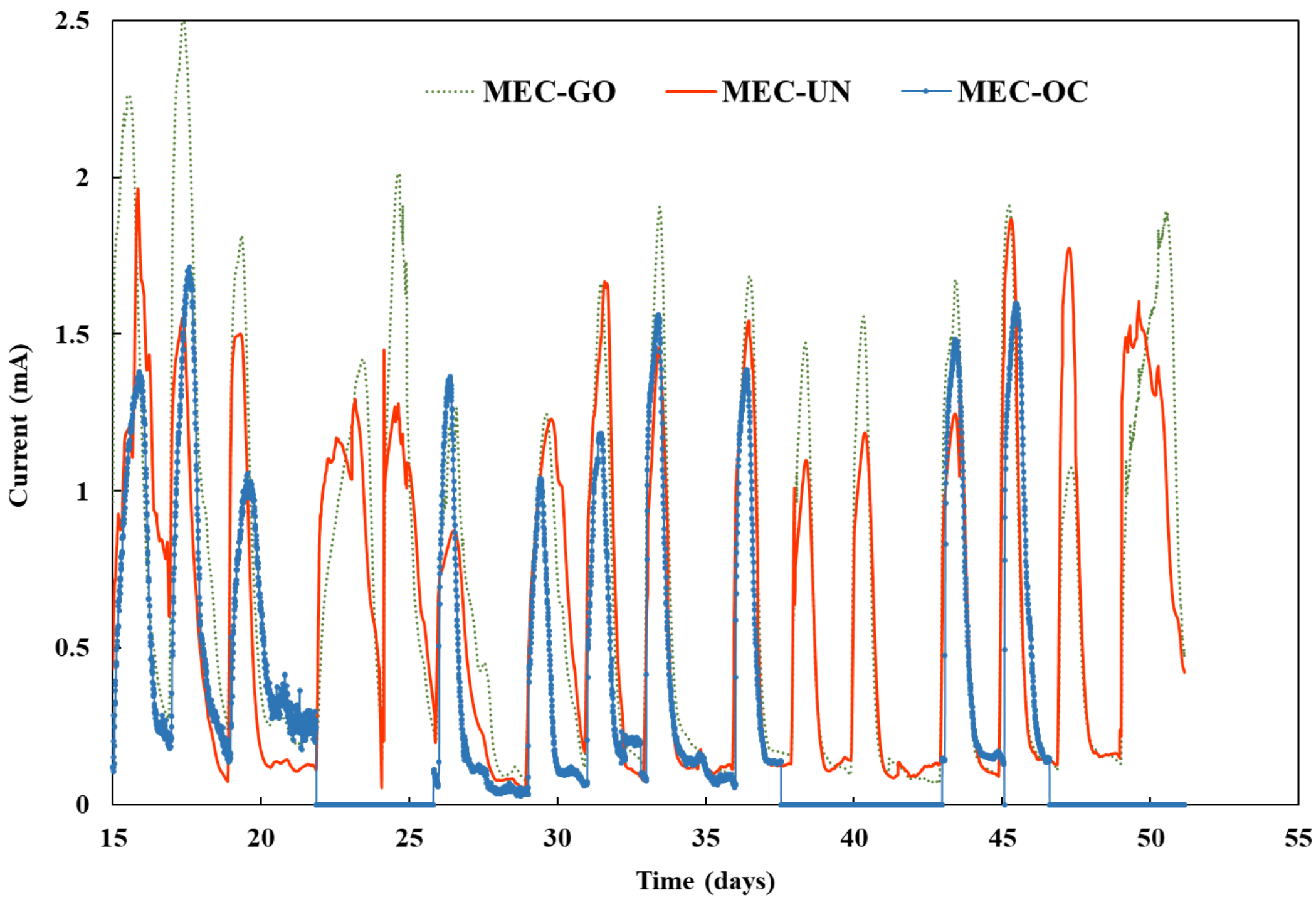

**Figure 1. Averaged current profiles obtained over the 36 days of operation for the three conditions investigated: graphene oxide-electrodeposited anodes (MEC-GO), MECs operated periodically in an open circuit (MEC-OC), and the control MECs (MEC-UN). The black squares mark the periods in which MEC-OC was operated in the open circuit.**

During the last batch cycle (starting at day 49), samples were regularly taken to analyse the TOC removal from the synthetic medium (Figure 2), where acetate accounted for 99% of the TOC in the samples. In this cycle, the MEC-GO cell produced an electrical charge of 240 C (averaged), which represented an 8% increase compared to that of MEC-UN (220 C). This slight difference may also explain the slightly better TOC removal observed for MEC-GO compared to that of MEC-UN (73% vs. 67%). However, the 57% TOC removal found for MEC-OC (which was operated in open-circuit conditions during this cycle) confirmed that not all the organic matter removal could be attributed to electrogenic processes. To determine the fraction of the electrons from acetate that ended up in the electrical circuit, the coulombic efficiency (CE) was evaluated. The slight differences in TOC removal and current between MEC-GO and MEC-UN agree with the similar values of the CE (~60%) for both reactors. This result also suggests that the bacteria that are unable to utilize the electrode as an electron acceptor are likely to use acetate as substrate. Overall, the graphene-modified electrode appeared to only slightly improve (3%) the TOC removal compared to the untreated electrode and 14% compared to an open-circuit MEC.

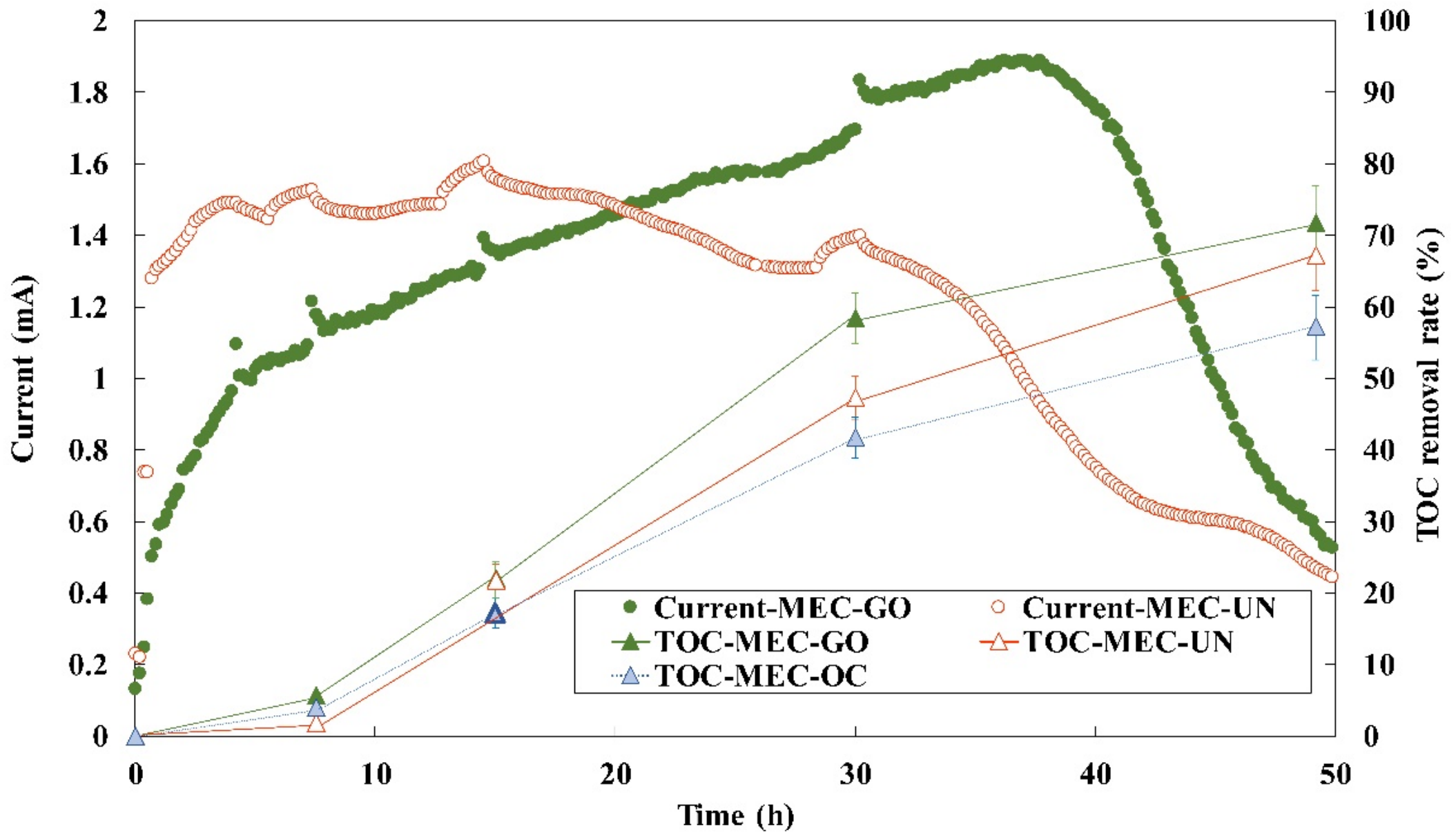

**Figure 2. Evolution of total organic carbon removal rate (%) and current (mA) during the last batch operation time.**

## 3.2. MBT degradation routes

Samples taken at 0, 7.5, 15, 30, and 50 h during the last batch cycle were analysed using LC-ESI-MS in positive and negative ionization modes to identify the main transformation products. The results are compiled in Table 1 and compared to the data available in the literature (Allaoui et al., 2010; B. Li et al., 2005; Borowska et al., 2016; Haroune et al., 2004; Malouki et al., 2004; Serdechnova et al., 2014; Zajíčková and Párkányi, 2009; Zajíčková and Párkányi, 2008). Only the main peaks were identified, and appropriate structures were proposed by considering the observed (m/z) ratios. The observed results are summarized in Table A1, in the supporting information.

Based on the observed products, a mechanistic map has been proposed that includes nine different routes, as shown in Figure 3. The observed products, considering the chemical structure of MBT, are not compatible with chemical oxidation or hydrolysis.

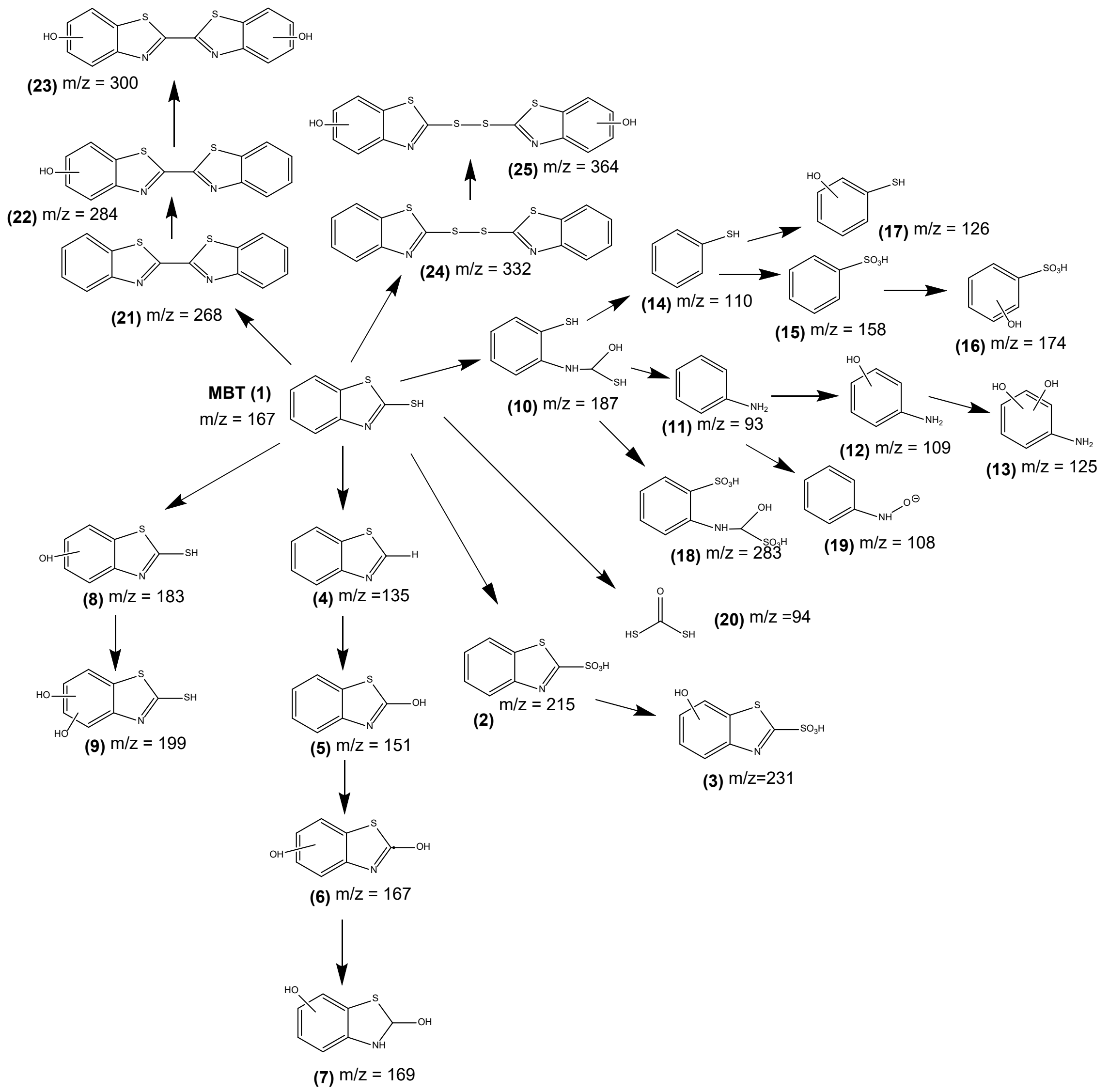

**Figure 3. Mechanistic routes proposed for the transformation of MBT under the reaction conditions**

We found evidence of at least five different degradation routes and two additional routes that lead to dimerization of the starting material. Oxidation of the thiol group to a sulfonic acid leads to **2**, which yields **3** upon $HO^{\bullet}$ additions to the aromatic ring. C-$S_{thiol}$ fragmentation (possibly homolytic) in MBT (**1**) followed by reaction with water leads to **4** and **5**, which render **6** and **7** through successive $HO^{\bullet}$ additions to the aromatic moiety. Similarly, consecutive additions of $HO^{\bullet}$ to the aromatic ring of MBT yield **8** and **9**. Homolysis of C-$S_{ring}$ followed by reaction with water gives **10**, from which three additional pathways open: i) C-N and C-S fragmentations to give **11**, which may then undergo $HO^{\bullet}$ addition to **12** and **13** or N-oxidation to **19**, ii) C-N fragmentation to **14**, which can then undergo oxidation to **15** followed by $HO^{\bullet}$ addition to **16** or direct $HO^{\bullet}$ addition to **14** to yield **17**, and, finally, iii) oxidation of both thiol groups to **18**. The finding of product **20** shows that there must be an alternative pathway through C-S and N=C fragmentation, which would also render **11** as a reaction product. C-$S_{thiol}$ bond hemolysis yields a C-centred radical that gives C-C dimerization to **21**, which can undergo successive $HO^{\bullet}$ additions to make **22** and **23**. Finally, the thiol can undergo reduction to disulfur **24**, from which, after two HO$^{\bullet}$ additions, **25** was obtained (a single hydroxylation intermediate was expected to exist, but we could not identify it among the reaction products). The observed mechanistic routes are in good agreement with previous work (Redouane-Salah et al., 2018).

## 3.3. Microbial community analysis on the anode

To gain further insight into the role of microorganisms on the performance of MECs and on the MBT degradation routes, the anodic microbial communities of MEC-GO, MEC-UN and MEC-OC were analysed using the 16S rRNA gene sequences approach and were studied at the phylum, family, and genera levels. At the phylum level (Fig. 4), the result showed no large differences between the anodes of the MECs subjected to the voltage (MEC-GO and MEC-UN) with *Proteobacteria* and *Bacteroidetes* being the dominant phyla (both with known hydrolysing and electrogenic capabilities (Zakaria et al., 2019)(Zhang et al., 2011)). *Firmicutes*, which are also capable of exocellular electron transfer (Zhao et al., 2018a), was the third dominant phylum in both (MEC-UN and MEC-GO). All of them, *Proteobacteria, Bacteroidetes*, and *Firmicutes*, are frequently found in the anodic microbial communities of BESs (Zhao et al., 2018b)(Hassan et al., 2018).

Despite the referred similarities, and while *Geobacteraceae, Campylobacteraceae*, and *Commamonadaceae* were common in both reactors, other families, such as *Hydrogenophilaceae, Rhodospirillaceae*, and *Rhodocyclaceae*, were only detected in the MEC-GO. These results are in agreement with our previous observation that graphene-modified electrodes promote a diversity of microbial communities compared to untreated electrodes (Alonso et al., 2017).

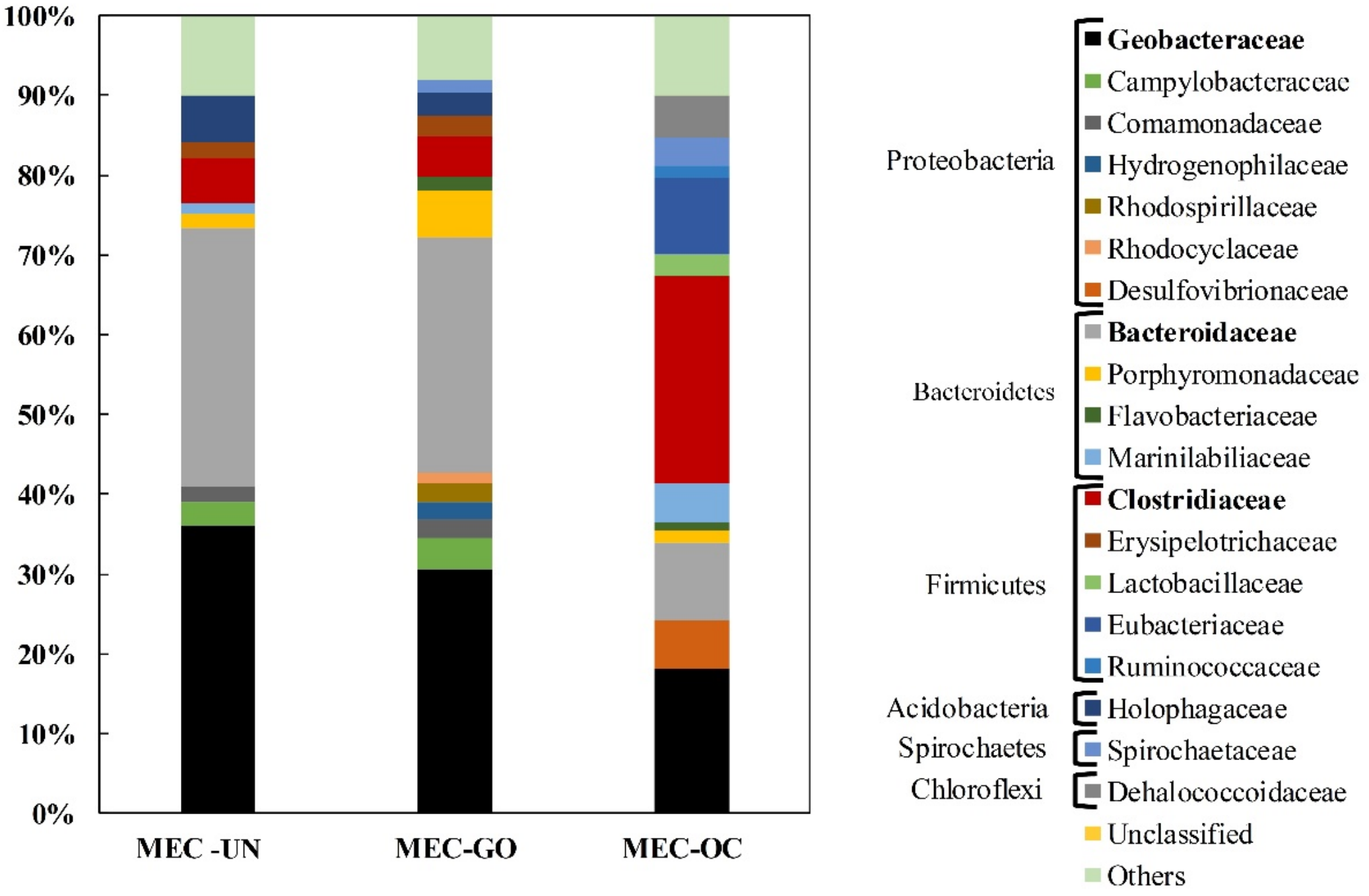

**Figure 4. Taxonomic classification of eubacterial communities at the family levels and the phyla to which these families belong for MEC-UN, MEC-GO, and MEC-OC**

In the case of the cell that was operated periodically under open-circuit conditions (MEC-OC), the anode developed a very different microbial environment with *Firmicutes* being the dominant phylum. Remarkably, we found the presence of the *Clostridiaceae* family (26%), which is known to include a variety of fermentative bacteria that decompose organic macromolecules into small organic acids, alcohols, and hydrogen (Miyahara et al., 2013), and the *Eubacteriaceae* (10%), which was not found in any of the other reactors. The latter is capable of enhancing more complete oxidation of substrates (Lei et al., 2018); thus, we hypothesize that this family may have been responsible for some of the TOC reduction when the MEC-OC was operated in the open circuit. Finally, despite the decrease in the relative abundance, the *Geobacteraceae* and *Bacteroidaceae* remained present.

Heat map analysis (Fig. 5) provides a visual way to compare the differences in microbial community structures at the genera level. *Geobacter*, which plays an important role in electricity generation by transferring electrons to the anode electrode (Karluvalı et al., 2015), was much more enriched at MEC-UN (36%) and MEC-GO (31%) than at MEC-OC (18%). This large difference between their relative abundances can only be attributable to the power outages in MEC-OC. Regardless, *Geobacter* was present in all of them because acetate was used as the organic carbon source to grow the exoelectrogenic biofilm (Yates et al., 2012). A similar trend was found for *Bacteriodes* (Figure 5), which has been reported to be capable of hydrolysing complex organics (Rismani-Yazdi et al., 2013). The larger abundance of bacteria capable of degrading complex organic matter (*Bacteriodes*) combined with the larger abundance of electrogenic bacteria (*Geobacter*) can likely explain the disparity in the degradation mechanism of MBT between the MECs where voltage was constantly applied (MEC-UN and MEC-GO) with respect to the MECs subjected to power outages (MEC-OC).

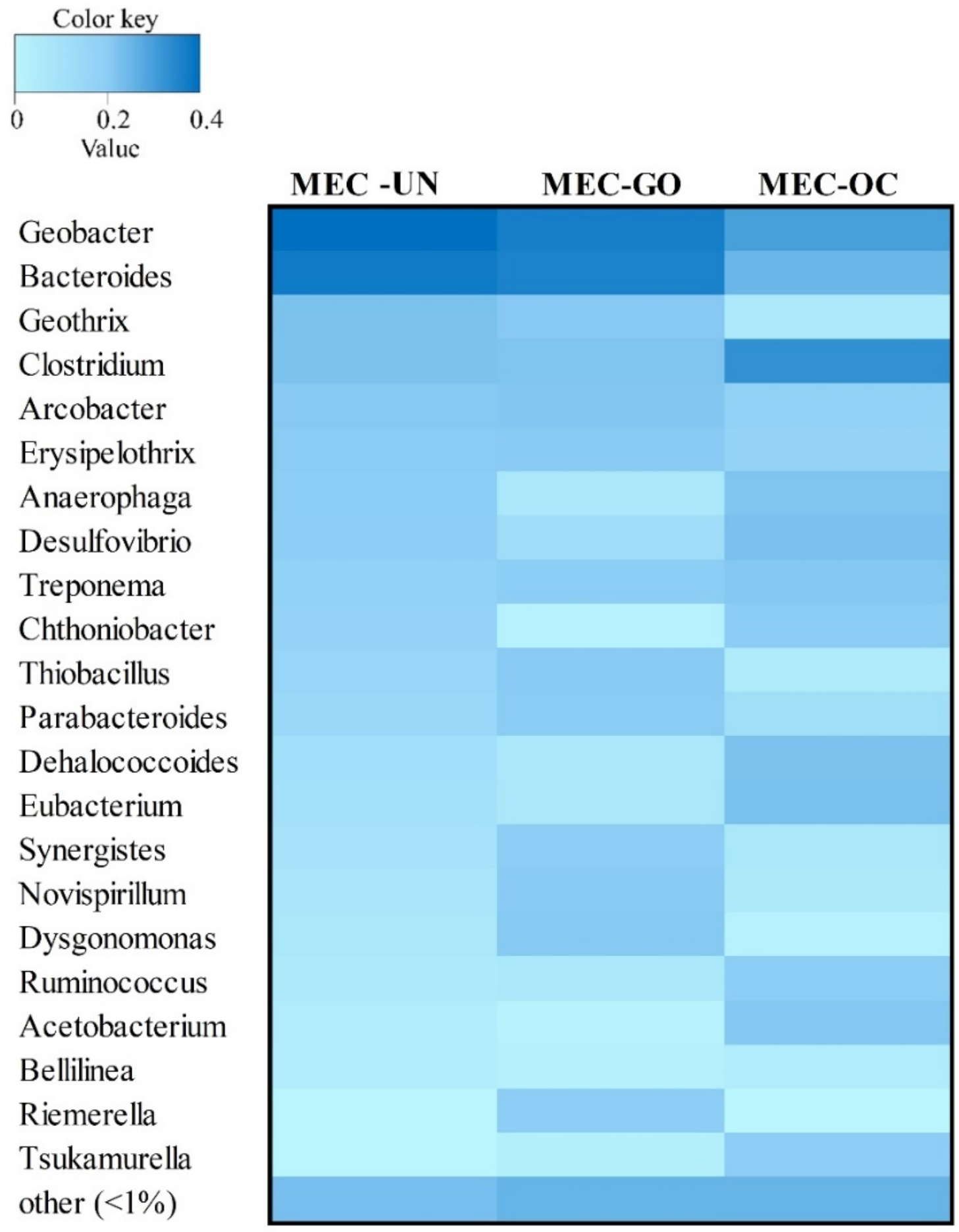

**Figure 5. Heat map summarizing the main genera present at the MEC-UN, MEC-GO, and MEC-OC biofilms**

To further confirm the experimental observations presented above, microbial characterization of the biomass was carried out using qPCR. This technique showed a significantly greater presence of biomass in the anodes of MEC-GO (9.6 x $10^7$ gene copy number · $g_{dw}^{-1}$ anode) and MEC-UN (4.4 x $10^7$ gene copy number· $g_{dw}^{-1}$ anode) compared to that of MEC-OC (1.3 x $10^6$ gene copy number · $g_{dw}^{-1}$ anode), which can only be attributed to the electrical outages in MEC-OC. It is also possible that this higher biomass in MEC-UN and especially in MEC-GO accelerates the MBT degradation owing to the presence of a greater quantity of functional bacteria.

# 3.4 MBT microbial degradation and biotoxicity

Very similar chromatograms were obtained from the effluents at different operational conditions. However, consecutive additions of OH substituents to the benzene ring of MBT, which lead to compounds **8** and **9** (Figure 3), was only detected in the MEC-GO effluent. We hypothesize that *Rhodococcus rhodochrous* was the main responsible for this, although due to its low relative abundance (<1%) its family (*Nocardiaceae*) and its genera (*Rhodococcus*) were not shown in Figure 4 and 5, respectively. The formation of 6-OH-MBT by *Rhodococcus rhodochrous* was previously described by Haroune et al. (2004), who also reported that hydroxylation of benzene ring could be formed by action of monooxygenase and hydroxylated MBT was less toxic than MBT. Moreover, the biotoxicity tests showed that while the MEC-OC reduced the biotoxicity of the effluent from an initial value of 46.2 eqtox·$m^{-3}$ to 44.8 eqtox·$m^{-3}$ (3% reduction), the MEC-UN reduced it to 39.7 eqtox·$m^{-3}$ (14% reduction), and the MEC-GO further reduced it to 27.9 eqtox·$m^{-3}$ (39% reduction). Therefore, the applied voltage and, primarily, the graphene electrodeposition had a significant impact on the reduction of the harmful effects on life of the MBT. It can be hypothesised, then, that the higher microbial diversity, combined with the presence of *Rhodococcus rhodochrous* could explain the greater reduction in biotoxicity observed in the MEC-GO.

# 4. CONCLUSIONS

The results presented in this study suggest that the anode of a MEC can significantly reduce the biotoxicity of MBT-contaminated streams. This may be related to the higher amount of biomass and to the larger abundance of the *Bacteroides* (able to degrade complex substrates) and *Geobacter* genera in the cells where voltage was constantly applied. The results also show that graphene-modified anodes can further reduce the biotoxicity, which seems linked to the greater diversity promoted by these anodes and especially to the presence of *Rhodococcus rhodochrous*. This bacterium can convert MBT into hydroxylated MBT (less toxic than MBT), which is in agreement with the MBT degradation mechanism described in this paper. Finally, the performance of the BES, in terms of current production, coulombic efficiency, and COT removal, was not significantly affected by the presence of graphene in the anodes or by periodically operating the BES under open-circuit conditions.

## Acknowledgments

This research was possible thanks to the financial support by 'Consejería de Educación de la Junta de Castilla y León' (ref: LE320P18), a project co-financed by FEDER funds. R. M. Alonso thanks the University of León for the predoctoral contract. M. Canle acknowledges financial support from the Ministerio de Economía y Competitividad (Spain) through project CTQ2015-71238-R (MINECO/FEDER), and regional government Xunta de Galicia (Project GPC ED431B 2017/59), respectively.